\documentstyle[multicol,aps,prl]{revtex}
\renewcommand{\narrowtext}{\begin{multicols}{2} \global\columnwidth20.5pc}
\renewcommand{\widetext}{\end{multicols}\global\columnwidth42.5pc}

\begin{document}

\title{Coulomb drag of Luttinger liquids and quantum-Hall edges}

\author{Karsten Flensberg}

\address{Danish Institute of Fundamental Metrology,\\
Building 307, Anker Engelunds Vej 1, DK-2800 Lyngby, Denmark}

\date{February 19, 1998}

\maketitle

\begin{abstract}

We study the transconductance for two coupled one-dimensional wires or edge
states described by Luttinger liquid models. The wires are assumed to
interact over a finite segment. We find for the interaction parameter
$g=1/2$ that the drag rate is finite at zero temperature, which cannot
occur in a Fermi-liquid system. The zero temperature drag is, however, cut off
at low temperature due to the finite length of the wires. 
We also consider edge states in the fractional quantum Hall regime, 
and we suggest that the low temperature enhancement of the drag effect 
might be seen in the fractional quantum Hall regime.

\end{abstract}


\narrowtext


One-dimensional (1D) systems have attracted much attention since the 
advances in lithographical fabrication techniques have made it possible to study
systems such as e.g. quantum point contacts, quantum wires,
quantum dots, and nano-tubes. Interacting 1D systems are
particularly interesting, because they are believed to be Luttinger liquids (LLs)
and thus exhibit non-Fermi liquid behavior. However, it is still not clear to what extent the
non-Fermi liquid behavior can be seen in a transport experiment
for clean systems, {\it i.e.} without impurity scattering. Several
authors\cite{finitelutt} have shown that the interactions do not influence
the conductance, and the reason for a possible interaction induced
effect observed\cite{exp} at finite temperature remains unexplained.

Another very interesting testing ground for LL behavior is
edge states in the  fractional quantum Hall (FQH) regime, 
where edge states can be
described as chiral LLs\cite{wen92}. Surprisingly, experiments have
shown that the tunneling between two edge states follows the LL
behavior\cite{gray98} both for compressible and incompressible states. 
Theoretical descriptions in terms of 1D edge channels have
been developed\cite{alei95glaz1}, and several authors\cite{1Dcomp,khve97} 
have recently extended this idea and calculated tunneling density of states 
which offers an explanation of the observed characteristics.

In this paper, another experiment which measures the
interaction effects is suggested, namely a Coulomb drag experiment. Coulomb
drag has during recent years proved  to be a powerful tool for
studying interaction and screening properties of coupled two-dimensional
electron systems\cite{plasmondrag}, including phonon-mediated
interactions\cite{phonondrag}, coupled QH systems both in the
integer regime\cite{IQHEdrag} and in the fractional regime where recent
experiments show Coulomb drag that saturates at the lowest temperatures at
filling factor close to one half\cite{lill98}. This type of behavior cannot be
explained within the present weak coupling theories for bulk composite-Fermion 
drag\cite{cf}.

1D drag between two Fermi liquids has been studied in the
linear and the non-linear regimes in Ref.\ \onlinecite{hu96flenb}. In Ref.
\onlinecite{naza97} the non-linear transresistance for two LLs
at zero temperature was studied and it was argued that
absolute drag (equal currents in the two wires) is possible. 
Here we show a related effect  for the linear conductance.

We concentrate on the temperature dependence of the linear response. 
Utilizing a mapping to two decoupled LLs,
it is shown that a LL description shows zero temperature drag for
the case where the coupling parameter $g=1/2$ \cite{close}. Moreover, we find 
an interesting dependence of the length of the interaction region and a 
regime for $g<1/2$, 
where the drag current is almost equal to the drive current. 
Our results are applied to edge states in coupled FQH systems 
with a narrow contriction, which is 
different from the situation of the Ref. \ \onlinecite{lill98}.
\begin{figure}
 \vbox to 3cm {\vss\hbox to 4cm
 {\hss\
   {\includegraphics{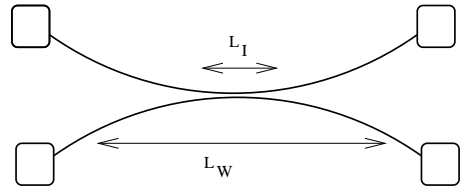}
   }
  \hss}
}
\caption{
Schematic outline of the geometry considered here. The two wires
are of length $L_W$, but they interact only in a region of length
$L_I$.}
\end{figure}


In our model the two spin-less LLs are coupled by Coulomb interactions, and
interwire tunneling is neglected. They are coupled in a finite region of
length $L_I$. The wires have lengths $L_W$ and it is assumed that the intrawire 
interaction is constant in this region, see Fig.~1. Thus the results are valid for
temperatures or voltages larger than the cut-off given by the energy of  charge
excitations with wavelengths of order $L_W$, i.e., $T>T_W=\hbar v_F/L_W$,
where $v_F$ is a Fermi velocity. 

The Hamiltonians for the separate systems are those of two
LLs ($i=1,2$) (we use $\hbar=k_B=1$)
\begin{equation}
\label{Hi}
              H_{0i} = \frac{v_{i}}{2}\int dx\,\left([P_i(x)]^2+\frac{1}{g_i^2}
              [\partial_x \phi_i(x)]^2\right),
\end{equation}
where $g_i,v_i$ are the interaction parameters
and Fermi velocities, respectively, and where
the densities, $\rho_L$ and $\rho_R$, of left and right movers
respectively, enter as
$\partial_x \phi_i(x) = [\rho_{Li}(x)+\rho_{Ri}(x)]\sqrt{\pi}$
and $P(x) = -[\rho_{Ri}(x)-\rho_{Li}(x)]\sqrt{\pi}$. The
fields $\phi_i$ and $P_i$ 
form conjugate variables: $[\phi_i(x),P_j(x')]=i\delta_{ij} \delta(x-x')$.

The Coulomb interaction between the two wires is given by
\begin{equation}
             H_{\mathrm{int}} = \int dx dy\,  U_{12}(x,y)
\rho_1(x)\rho_2(y).
\end{equation}
Note that the interwire interaction $U_{12}(x,y)$ only acts within a region of
length $L_I$.
Here the subsystem densities are given by
\begin{equation}
\label{rhodef}
\rho_i(x) = \rho_{Li}(x)+\rho_{Ri}(x)+\Psi_{Ri}^\dagger(x)\Psi_{Li}(x)+
\Psi_{Li}^\dagger(x)\Psi_{Ri}(x),
\end{equation}
where  $\Psi_{L(R)i}$ are Fermion operators corresponding to the
left(right) movers. In the bosonization language these are given by\cite{soly:rev}
(one set for each wire)$ \Psi_{L(R)}(x) = 
\exp\left(\pm i k_{F}x \mp 2\pi \int_{-\infty}^x dy\rho_{L(R)}(y)\right)/\sqrt{2\pi \alpha}$, 
$\alpha$ is a cut-off parameter\cite{phases}.

We separate the interwire interaction in 4 terms describing the
different possibilities of forward and/or backward scattering. Note that
since the interaction region is finite, momentum need not be conserved
in the scattering process. We have
\begin{eqnarray}
H_{\mathrm{int}} &=& \int dx dy\, U_{12}(x,y)\times \nonumber\\
&& \left[B^a(x,y)+B^b(x,y)+B^c(x,y)+B^d(x,y)\right],
\end{eqnarray}
where
\begin{mathletters}
\begin{eqnarray}
B^a(x,y) &=&  A_1(x)A_2^\dagger(y)+\mathrm{h.c.}, \\
B^b(x,y) &=&  A_1(x)A_2(y)+\mathrm{h.c.}, \\
B^c(x,y) &=& \left[\partial_x \phi_1(x) \partial_y
\phi_2(y)\right]/\pi  , \\
B^d(x,y) &=&  \phi_1(x)\left[A_2(y)+A_2^\dagger(y)\right]/\sqrt{\pi}
+1 \leftrightarrow 2.
\end{eqnarray}
\end{mathletters}
Here $A(x)$ is a "backscattering operator" defined as $
A(x) = \exp\left(i2k_Fx +i2\sqrt{\pi} \phi(x) \right)/(2\pi \alpha)$.

The two terms $B^b$ and $B^d$ correspond to non-momentum-conserving
scatterings processes. The terms $B^c$ and $B^d$ do not provide a mechanism
for Coulomb drag (to any order in perturbation theory) and, furthermore,
since the renormalization due to these terms is not important at low
energies, they are omitted. We will make one further approximation which is
valid when the relevant energy scale is  smaller than $v_F/L_I$.
In this limit, the spatial dependence of $\phi$'s in the backscattering operator
can be neglected. The final expression for the interaction now reads
\begin{eqnarray}\label{Hintfinal}
H_{\mathrm{int}} &=& \frac{1}{2\pi^2\alpha^2}\int dx dy\,
U_{12}(x,y)\times \nonumber\\&&
\left\{\cos[2(k_{F1}x-k_{F2}y)+2\sqrt{\pi}(\phi_1(0)-\phi_2(0))]\right.
\nonumber\\&&
\left.+\cos[2(k_{F1}x+k_{F2}y)+2\sqrt{\pi}(\phi_1(0)+\phi_2(0))]\right\}.
\end{eqnarray}


Now we transform the Hamiltonian to that of two (interacting)
LLs scattering against a single impurity potential.
Define the new field operators
\begin{mathletters}
\label{rotation}
\begin{eqnarray}
\Phi &=& \phi_1+\phi_2,\quad P = (P_1+P_2)/2,\\
\Theta &=& \phi_1-\phi_2,\quad \Pi = (P_1-P_2)/2,
\end{eqnarray}
\end{mathletters}
defined such that $\Phi,P$ and $\Theta,\Pi$ are conjugate pairs. 
The interwire interaction term becomes particularly simple in this basis and
the Hamiltonian transforms to 
\begin{equation} H = H_0+H'+H_{\mathrm{int}}, 
\end{equation}
where
\begin{eqnarray}
H_0 &=& \frac{\bar{v}}{2}\int dx
\left([P(x)]^2+\frac{1}{\bar{g}^2} [\partial_x \Phi(x)]^2\right)
\nonumber\\
&&+\left([\Pi(x)]^2+\frac{1}{\bar{g}^2}[\partial_x \Theta(x)]^2\right), 
\label{H0trans}
\end{eqnarray}
and where $H'$
describes the interaction between the new field operators
\begin{equation}
\label{Hm} H' = \bar{v}\int dx \left[a P(x)\Pi(x) +\frac{b}{\bar{g}^2}
\partial_x \Phi(x)\partial_x \Theta(x)\right],
\end{equation}
where
$\bar{v} = (v_1+v_2)$, $1/\bar{g}^2=(v_1/g_1^2+v_2/g_2^2)/4\bar{v}$,
$a=(v_1-v_2)/\bar{v}$, and $b =(v_1/g_1^2-v_2/g_2^2)/(v_1/g_1^2+v_2/g_2^2)$.


The current operator for the LL model is
given by $j_i=v_F P_i/\sqrt{\pi}$ and through the continuity equation,
the current is expressed as $j_j(x) = - \partial_t \phi_j(x,t)/\sqrt{\pi}$.
Using the Kubo formula, we obtain the transconductance in terms of the new
fields defined in Eq.\ (\ref{rotation}) as 
\begin{equation}
G_{21}(\omega) = \frac{i\omega e^2}{4\pi}\left[D_{\Phi}^r(x,x';\omega)
-D_{\Theta}^r(x,x';\omega)\right],
\end{equation}
where the Green's functions
\begin{mathletters}
\begin{eqnarray}
D_\Phi(t-t')&=&-i\Theta (t-t')\langle
[\Phi(x,t'),\Phi(x',t)]\rangle,\\
D_\Theta(t-t') &=&-i\Theta (t-t')\langle
[\Theta(x,t),\Theta(x',t')]\rangle,
\end{eqnarray}
\end{mathletters}
have been defined (note that $D(x,x';\omega)$ is independent of $x,x'$ in the dc-limit).


For identical wires the part of the Hamiltonian which couples the $\Phi$
and $\Theta$ sector in Eq.\ (\ref{Hm}) is equal to zero. The remaining
Hamiltonian is equivalent to two LLs models scattering
against single impurities, but with new interaction parameters.
We can therefore use the results from this well-studied problem\cite{kanefisher,fend95}.  
The transconductance simplifies to
\begin{equation}
\label{G21lutt}
G_{21} = \frac{1}{4}\left(G_{\mathrm{Lutt}}(V_1,2g)-G_{\mathrm{Lutt}}(V_2,2g)\right),
\end{equation}
where $g\equiv g_1=g_2$ and where $G_{\mathrm{Lutt}}(V,2g)$\cite{Glutt}
is the conductance of a LL with interaction parameter, $2g$, scattering
against a single impurity with a backscattering amplitude $V=V(2k_F)$. 
(Below it is shown that Eq.\ (\ref{G21lutt}) is valid even when the velocities are different.)
Here we have defined
\begin{equation}
V_{1,2} = \frac{D}{2\pi v_F}
\int dx dy\, U_{12}(x,y) \exp\left[2ik_F(x \pm y)\right],
\end{equation}
where the small momentum cut-off is parametrized in terms of a high
energy cut-off: $D = v_F/\alpha$.

Now several conclusions follow. Firstly, it is seen that for a short interaction
region $L_Ik_F \ll 1$, which implies $V_1 \approx V_2$, all momentum transfers 
have equal weights and hence there is no Coulomb drag\cite{g11}.
More importantly, the scaling properties of the model can be 
read out\cite{kanefisher,fend95}: 

For $g>1/2$, $G_{\mathrm{Lutt}}$ goes to a constant as $T\rightarrow 0$, which 
means that  $G_{21}$ goes to zero at zero temperature. The corrections to
the low temperature limit give the power law:
$G_{21} \sim T^{4 g-2}$. For non-interacting wires the drag
effect is thus quadratic in temperature; in contrast to the case
where they interact throughout the wires, in which case the drag scales
linearly with temperature\cite{hu96flenb}.

For $g<1/2$, both terms in Eq.\ (\ref{G21lutt}) go to zero, because the
model scales toward strong backscattering. The power laws are however the
same but the prefactors will be different and we can conclude that $G_{21}
\sim T^{1/g-2}$.

For $g=1/2$ the problem maps to Fermi-liquids and we get a temperature
independent $G_{21}$ and hence only for $g=1/2$ does the
transconductance remain finite as temperature goes to zero.


Consider again the case of identical wires in the vicinity of $g=1/2$, where the
problem maps to that of two Fermi-liquids. We may use the perturbative
renormalization method developed in Ref.\ \onlinecite{yue94} rather than
the exact solution based on the Bethe ansatz\cite{fend95}, and we obtain for
the transconductance\cite{Glutt}
\begin{equation}
\label{G21nonpert}
G_{21}= \frac{e^2}{4\pi} \frac{(|W_2|^2-|W_1|^2) t^\gamma}
{[1+|W_1|^2 t^\gamma][1+|W_2|^2  t^\gamma]},
\end{equation}
where $t=(T/D)$, $W_i=V_i/v_F$, and $\gamma=4g-2$.

In Fig. 2, we show $G_{21}$ expressed in Eq. (15) for different 
parameters.
The interwire interaction $U_{12}$ is calculated for
typical parameters for GaAs quantum wires and the distance between 
the wires is chosen to be $2/k_F$.
In accordance with the arguments given above, the transconductance 
peaks when $g=1/2$ at low temperatures and the peak moves to smaller
$g$ values for higher temperatures. (Furthermore, 
the stronger the interwire interaction the closer is the peak 
position to $g=1/2$.) 
Therefore for $g<1/2$, $G_{21}$ shows non-monotonic
behavior as a function of temperature.

For a long interaction region, i.e. $W_1\ll 1$,
$G_{21}$ approaches the value $e^2/2h$ for small temperatures 
(but still larger than $T^* = D |W_1|^{-2/\gamma}$), 
see inset of Fig. 2. 
In this regime, {\em the diagonal 
conductance\cite{g11} also tends to $e^2/2h$ and thus the currents in
two wires become the same}, which is similar to the absolute 
drag effects found in Ref.\ \onlinecite{naza97}.
This interesting effect occurs because the momentum conserving 
backscattering increases with 
decreasing temperature (second term of Eq.\ (\ref{G21lutt}) 
goes to zero). It saturates when the two currents 
are the same and the net momentum exchange hence is zero.
\begin{figure}
 \vbox to 5 cm {\vss\hbox to 5 cm
 {\hss\
   {\includegraphics{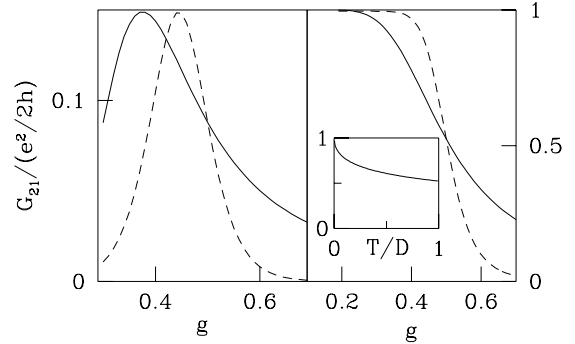}
   }
  \hss}
}
\caption{
Transconductance for two identical wires using the expression derived for
$g$ close to one half. The full (dashed) curves are for $T/D =$ 0.1
(0.001). For a short interaction region (left panel, where $k_FL_I=1$)
the transconductance peaks near $g=1/2$. The peak moves toward
$g=1/2$ for smaller temperature. For long interaction region 
(right panel, where $k_FL_I=20$), the peak is not
developed and instead $G_{21}$ approaches $e^2/2h$ at low
temperature for $g<1/2$, which is shown in the inset for 
$g=0.4$. This regime corresponds to the limit 
where only momentum conserving scattering is possible and where the
currents in two wires become almost the same. } 
\end{figure}
 

In the general case, where the wires have different $g$-values and
velocities, there is a coupling term in the transformed Hamiltonian, Eq.\
(\ref{Hm}). Since the conductance involves only the fields at $x=0$, we 
integrate out all the $x\neq 0$ fields. The resulting action reads
\begin{mathletters}
\label{Seff}
\begin{eqnarray}
&&S=S_0+S_{int},\\
&& S_0 = \frac{1}{\beta }\sum_{\omega_n } \frac{|\omega_n|}{\bar{g}}
\left(\Phi_{\omega}\; \Theta_{\omega} \right)^*
\left(\begin{array}{cc}
k_+ & k_{-}\\
k_{-} & k_+\\
\end{array}\right)
\left(
\begin{array}{c}
\Phi_{\omega}\\
\Theta_{\omega}
\end{array}\right),\\
&&S_{int} = \int_{0}^{\beta}d\tau \left\{ V_2 e^{2i\sqrt{\pi}\Theta(\tau)}
+  V_1e^{ 2i\sqrt{\pi}\Phi (\tau)} + \mathrm{c.c.}\right\},
\end{eqnarray}
\end{mathletters}
where
$ k_\pm=\frac{1}{2}\left[(v_2+v_1)(g_1\pm g_2)^2/(g_1^2v_2+g_2^2v_1) \right]^{1/2}$.
For the special case, 
when the two wires have the same interaction parameter (but unequal 
velocities) the two sectors separate and the solution is again
given by Eq.\ (\ref{G21lutt}). For the general situation,
we perform a perturbative renormalization 
group calculation and find
\begin{equation} \label{Renor}
\frac{d V_i}{d\ln D} = (1-(g_1+g_2)) V_i.
\end{equation}
The flow is marginal when the sum of the $g$-constants 
is equal to one and
the zero temperature drag predicted above occurs at this line and for
$g_1+g_2< 1$ a strong enhancement of the Coulomb drag occurs.


Finally, we consider Coulomb drag of edge excitations in the FQH
regime. Recently, there has been a large activity trying to 
understand the low energy edge excitations and the tunneling 
experiments\cite{gray98} in FQH states\cite{1Dcomp,khve97}. 
These works show that the edge excitations can be described
as unidirectional bosonic edge excitations, which
were then used to form an electron charge creation
operator as input in an independent boson model for the tunneling density
of states. 

Following these works, we write the Hamiltonian
that governs the dynamics for (say) the left edge branch as 
$H_{\mathrm{edge,L}} = \frac{2\pi}{L} v_D \sum_{k>0} \rho_L(k)\rho_L(-k),$
where $\rho_L$ is the left branch 1D charge operator and 
$[\rho_L(-k),\rho_L(k)] = \nu k L/2\pi$. Here
$\nu$ is the filling factor, $v_D$ is the drift velocity at the edge 
and $L$ a normalization length. Similar expression is obtained for the
right branch, similar to the Tomonaga-Luttinger model.
The operator which creates a  single charge $e$ at point $x$,
moving with velocity $v_D$, in the left channel is
$\Psi_L^\dagger \sim e^{-i k_D x}
\exp\left(\frac{2\pi}{\nu L}\sum_{k}\rho_L(k)e^{ik x}/k\right)$.
Notice that this operator does not fulfill the correct anticommutation
relations for fermions. However, the backscattering operator 
$A=\Psi_R^\dagger \Psi_L$ does obey the correct commutation relations
and below it is utilized to describe interedge tunnelings in a model
Hamiltonian for coupled FQH systems. 

The following experiment is proposed:  the  coupled QH systems are narrowed 
by a constriction such that edge states moving in opposite directions are coupled and
backscattering can occur. This system  can be modeled by writing the
backscattering interaction in terms of the backscattering operator, $A$. After some
simple algebra, we 
arrive at the Hamiltonian for two coupled LLs, Eqs.\ (\ref{Hi}) and (\ref{Hintfinal}) ,
with $g=\nu$ \cite{nonunig}
and conclusions similar to above therefore immediately follow from Eq.\ (\ref{G21lutt}): 
{\em In particular, we predict  non-zero
drag at zero temperature for half-filled Landau levels and furthermore near its maxium
value the drag effect increases with decreasing temperature (or voltage). 
For long interaction region, the transconductance goes to a  
universal value $\nu e^2/2 h$ at low temperature for $\nu < 1/2$.} 

In conclusion, we have calculated Coulomb drag for LLs and found 
interesting behavior near g=1/2 such as zero temperature drag and 
non-monotonic temperature dependence. These predictions can be 
tested for edge states in FQH systems. 

The author acknowledges valuable discussions with 
Ben Hu and Antti-Pekka Jauho. 

Recently, a related paper appeared\cite{komn98}. These authors discuss the
conductance of crossed LLs coupled at a single point 
and find very similar results to ours. 


\widetext

\begin{thebibliography}{10}
\bibitem{phases}
The Fermion operator should include a phase factor which 
ensures that the electron operators
corresponding to different branches anticommute. 
However, they are not important for the present problem.
\bibitem{Glutt} Be aware that in absence of scattering 
$G_{\mathrm{Lutt}}(2g) = 2e^2/h$, because  in this case $g$
should be set to one[1]. This is not the case for the FQH edge states.
\bibitem{g11}
The diagonal conductance is $G_{11}$ given by Eq. (\protect\ref{G21lutt}) 
with a plus instead of the minus and $G_{11}$ of course remains finite
for $V_1=V_2$.
\end{thebibliography}
\end{document}